\documentclass[superscriptaddress,nofootinbib,amsmath,amssymb,aps,prr,twocolumn]{revtex4-2}

\usepackage{graphicx}
\usepackage{dcolumn}
\usepackage{bm}
\usepackage{latexsym}
\usepackage{amsmath, dsfont, physics}
\usepackage{amssymb}
\usepackage{graphicx}
\usepackage{caption}
\usepackage{subfigure}
\usepackage{float}
\usepackage{mathrsfs}
\usepackage{color}
\usepackage{txfonts}
\usepackage[justification=centering,
            format=plain]{caption}
\renewcommand{\raggedright}{\leftskip=0pt \rightskip=0pt plus 0cm}

\begin{document}

\title{Topological transitions in dissipatively coupled Su-Schrieffer-Heeger models}

\author{Jayakrishnan M. P. Nair}
\email{jayakrishnan00213@tamu.edu}
\affiliation{Institute for Quantum Science and Engineering, Texas A$\&$M University, College Station, TX 77843, USA}
\affiliation{Department of Physics and Astronomy, Texas A$\&$M University, College Station, TX 77843, USA}

\author{Marlan O. Scully}
\email{scully@tamu.edu}
\affiliation{Institute for Quantum Science and Engineering, Texas A$\&$M University, College Station, TX 77843, USA}
\affiliation{Department of Physics and Astronomy, Texas A$\&$M University, College Station, TX 77843, USA}
\affiliation{Baylor University, Waco, TX 76704, USA}
\affiliation{Princeton University, Princeton, New Jersey 08544, USA}

\author{Girish S. Agarwal}
\email{Girish.Agarwal@ag.tamu.edu }
\affiliation{Institute for Quantum Science and Engineering, Texas A$\&$M University, College Station, TX 77843, USA}
\affiliation{Department of Physics and Astronomy, Texas A$\&$M University, College Station, TX 77843, USA}
\date{\today}

\begin{abstract}
Non-Hermitian topological phenomena have gained much interest among physicists in recent years. In this paper, we expound on the physics of dissipatively coupled Su-Schrieffer-Heeger (SSH) lattices, specifically in systems with bosonic and electrical constituents. In the context of electrical circuits, we demonstrate that a series of resistively coupled LCR circuits mimics the topology of a dissipatively coupled SSH model. In addition, we foreground a scheme to construct dissipatively coupled SSH lattices involving a set of non-interacting bosonic oscillators weakly coupled to engineered reservoirs of modes possessing substantially small lifetimes when compared to other system timescales. Further, by activating the coherent coupling between bosonic oscillators, we elucidate the emergence of non-reciprocal dissipative coupling which can be controlled by the phase of the coherent interaction strength precipitating in phase-dependent topological transitions and skin effect. Our analyses are generic, apropos of a large class of systems involving, for instance, optical and microwave settings, while the circuit implementation represents the most straightforward of them.
 
\end{abstract}

\maketitle

\section{Introduction}
One of the prime objectives of research in condensed matter physics is the characterisation of matter phases earmarked by the (spontaneous breaking of) symmetries of the system under consideration. In this context, the discovery of the quantum Hall effect marked a stark shift in the understanding of phases and introduced the concept of topological order, spawning the field of topological insulators \cite{RevModPhys.82.3045, RevModPhys.83.1057, RevModPhys.91.015006}. This was subsequently realized on a variety of different platforms, including, photonic \cite{lu2014topological}, cold atomic systems \cite{RevModPhys.83.1523} and many more \cite{susstrunk2015observation, kane2014topological, paulose2015topological}. One of the key features of the topological classification of phases is the bulk boundary correspondence (BBC) and the emergence of edge and surface states that are impervious to environmental loss and disorder with applications ranging from the realization of topological qubits \cite{kitaev2001unpaired, PhysRevB.81.014505, alicea2011non, PhysRevLett.110.076401, you2014encoding} to lasing \cite{st2017lasing, bahari2017nonreciprocal, harari2018topological, bandres2018topological} among others \cite{PhysRevLett.119.023603, PhysRevLett.115.045303, PhysRevLett.118.073602, PhysRevLett.119.173901, PhysRevLett.124.023603, PhysRevLett.127.250402, barik2018topological, PhysRevB.101.165427}.

Until recently, the lion's share of research on the physics of topological systems involved Hermitian models. However, real physical systems interact with their environment resulting in open quantum dynamics \cite{agarwal2006quantum} and effective non-Hermitian Hamiltonians. In the last few years, the topology of non-Hermitian lattice systems has been a subject of intense research activity \cite{RevModPhys.93.015005, PhysRevX.8.031079}, unravelling some exciting new physics, for example, the breakdown of BBC and skin effect in systems possessing non-reciprocal couplings \cite{PhysRevLett.116.133903, xiong2018does, PhysRevLett.121.086803, PhysRevLett.121.026808, PhysRevB.99.245116, PhysRevB.99.201103, PhysRevB.102.205118, xiao2020non, PhysRevA.106.L061302}. A quintessential model in the study of topological physics is the Su-Schrieffer-Heeger (SSH) model  \cite{malkova2009observation, tan2014photonic, PhysRevLett.115.040402, PhysRevB.96.045417, lee2018topolectrical} and several non-Hermitian extensions of the model have been considered in literature \cite{PhysRevA.97.052115, zhao2021real, PhysRevB.97.045106, PhysRevB.105.L201113, PhysRevA.102.022404}. For instance, \cite{PhysRevB.97.045106, PhysRevB.105.L201113} considered $PT$-symmetric extensions of the SSH model which can be engineered by the incorporation gain into the system. However, the physics of lattice models with a purely dissipative form of coupling \cite{leefmans2022topological, PhysRevLett.130.153602} between the constituents is largely unplumbed. Dissipative coupling between two otherwise non-interacting systems emanates from their decay into common dissipative channels \cite{agarwal2012quantum} and it is worth noting that dissipative couplings are more prevalent in nature compared to their coherent counterparts. Such couplings have been investigated both theoretically and experimentally in a multitude of settings \cite{choi2018observation, PhysRevA.100.013812, PhysRevA.101.063814, zhang2020breaking, silver2021nonlinear}, for example, involving magnonic and photonic sub-systems  \cite{PhysRevLett.121.137203, PhysRevB.100.094415, PhysRevLett.123.127202, PhysRevLett.126.180401, PhysRevB.103.224401}. 

In this work, we focus on the physics of dissipatively coupled SSH models. In particular, we demonstrate two distinct experimentally realizable schemes involving bosonic and electrical subsystems. We show that a system of resistively coupled LCR resonators mimics the topology of dissipatively coupled SSH (DSSH) models. Subsequently, we illustrate that a lattice of otherwise non-interacting bosonic oscillators interacting with engineered reservoirs of modes having significantly large decay parameters compared to other system parameters can be described by an effective non-Hermitian Hamiltonian akin to the DSSH model. Furthermore, by triggering the coherent coupling between the oscillators, we outline the generation of non-reciprocal coupling in DSSH models featuring topological transitions that can be controlled by the phase of the coherent interaction strength.  Note, \textit{en passant}, the generality of our results grants an immediate experimental realization of the protocols discussed in the subsequent sections, especially in the microwave and optical domains.

This paper is organized as follows. In section \ref{sec1}, we revisit the SSH model with coherent couplings followed by a discussion of its dissipative counterpart in section \ref{sec2}. Subsequently, we delineate two independent schemes comprising of electrical and bosonic constituents for the realization of the DSSH model in section \ref{sec3}. In section \ref{sec4}, we foreground a protocol for the realization of DSSH model with non-reciprocal couplings through the application of a coherent form interaction between bosonic oscillators, translating into topological transitions and skin effect. Finally, we conclude our results in section \ref{sec5}.

\section{Key features of the SSH model}\label{sec1}
\begin{figure}
\captionsetup{justification=raggedright,singlelinecheck=false}
 \centering
   \includegraphics[scale=0.94]{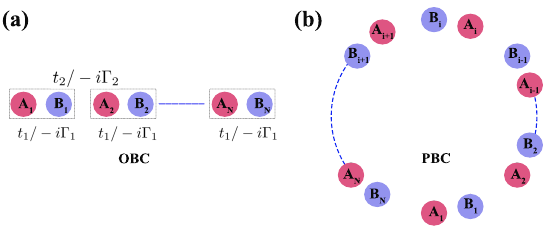}
\caption{Schematic of the system described by Eqs. ({\ref{eq1}-\ref{eq2}}) and Eq. (\ref{eq5.1}) under open and periodic boundary conditions.}
\label{sch}
\end{figure}

\begin{figure}
\captionsetup{justification=raggedright,singlelinecheck=false}
 \centering
   \includegraphics[scale=0.8]{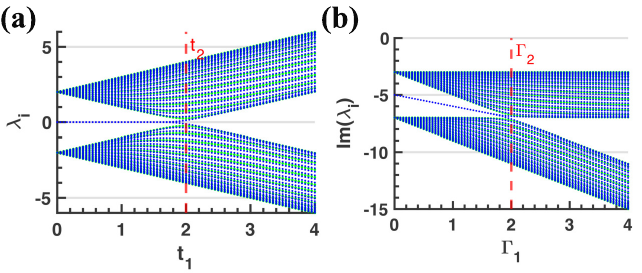}
\caption{(a) The eigenvalues of the Hermitian SSH model described by Eqs. (\ref{eq1}-\ref{eq2}) under periodic (green) and open (blue) boundary conditions; (b) The imaginary part of the eigenvalues of the dissipative SSH model described by Eq. (\ref{eq5.1}). under periodic (green) and open (blue) boundary conditions and the number of unit cells $N=25$ and the effective damping $\Gamma_r=\gamma+\Gamma_1+\Gamma_2$ and $\gamma=3$.}
\label{eigen}
\end{figure}

We begin by revisiting theSSH model coherently coupled unit cells. To this end, consider a one-dimensional (1-D) lattice of two different types of sites, A and B with staggered nearest neighbor couplings as depicted in Fig. \ref{sch}. The interaction Hamiltonian of the system subject to open boundary conditions (OBC) is given by
\begin{align}\label{eq1}
    H=\sum_{i=1}^{N}t_1\ket{A_i}\bra{B_i}+\sum_{i=1}^{N-1}t_2\ket{A_{{i+1}}}\bra{B_i}+h.c,
\end{align}
where $N$ denotes the number of unit cells, $\ket{A_i}$, $\ket{B_i}$ characterize the particle excitation at their respective location while $t_1$ and $t_2$ $\in\mathbb{R}$ are the intra and inter-cellular couplings respectively. Equivalently, we can write the SSH Hamiltonian subject to periodic boundary conditions (PBC) as
\begin{align}\label{eq2}
H=\sum_{i=1}^{N}t_1\ket{A_i}\bra{B_i}+\sum_{i=1}^{N}t_2\ket{A_{f(i)}}\bra{B_i}+h.c,
\end{align}
where $f(i)=i+1$ mod $N$. Invoking the Bloch theorem, the Hamiltonian under PBC can be recast in the Fourier domain in terms of the Bloch Hamiltonian provided by
\begin{align}\label{eq3}
H_k=\begin{pmatrix}
0&R(k)e^{-i\phi(k)}\\
R(k)e^{i\phi(k)}&0
\end{pmatrix},
\end{align}
where $k=\frac{2\pi n}{N}$, $m\in\{1,2..N\}$, $R(k)=\sqrt{t_1^2+t_2^2+2t_1t_2\cos{(k)}}$, the phase $\phi(k)=\arctan(\frac{t_2\sin{k}}{t_1+t_2\cos{k}})$ and we set the inter-cellular spacing $a=1$. Note that the Hermitian matrix $H_k$ is chiral symmetric, that is $\sigma_z{H}_{k}\sigma_z=-{H}_{k}$ precipitating in symmetric eigenvalues $E_{\pm}=\pm R(k)$ and corresponding eigenstates
\begin{align}\label{eq4}
    \ket{E_{\pm},k}=\frac{1}{\sqrt{2}}\begin{pmatrix}
\pm1\\
e^{i\phi(k)}
\end{pmatrix}.
\end{align}
Palpably, the gap between the energy eigenvalues vanishes at $k=\pi$ and $t_1=t_2$ as demonstrated in Fig. \ref{eigen} (a). In contrast, the Hamiltonian under OBC described by Eq. (\ref{eq1}) supports two zero-energy eigenvalues in the large $N$ limit for $\abs{\frac{t_1}{t_2}}<1$, eliciting the well-known edge modes of the SSH model, a testament to the non-trivial topology of the system. In addition, one can define a topological invariant, \textit{viz}, the winding number $\nu_\pm$ defined in terms of the Berry connection $A_{\pm}(k)=i\bra{E_\pm,k}\partial_k\ket{E_\pm,k}$ as 
\begin{align}\label{eq5}
    \nu_\pm=\frac{1}{\pi}\oint A_\pm(k)dk.
\end{align}
For instance, $\nu_+$ calculated from Eq. (\ref{eq4}) and Eq. (\ref{eq5}) satisfies
\begin{align}\label{eq6}
    \nu_+=\begin{cases}
  1 & \text{if $\abs{\frac{t_1}{t_2}}<1$} \\
  0 & \text{if $\abs{\frac{t_1}{t_2}}>1$}.
\end{cases}
\end{align}
A non-zero winding number $\nu_\pm$ is a direct manifestation of the non-trivial topology of the system demonstrating $2\abs{\nu_\pm}$ number of edge modes and $t_1=t_2$, the point of vanishing gap between bulk energy bands demarcates the boundary between the two phases. This is known as the bulk boundary correspondence (BBC) in Hermitian lattice systems. It is worth noting that coherent coupling between systems emanates from the spatial overlap between their respective modes. By contrast, dissipatively coupled systems with a non-Hermitian form of interaction are prevalent in nature. In essence, any two systems interacting with a common intermediary channel will spawn a dissipative form coupling. In the following, we will discuss the general properties of dissipatively coupled SSH models.
\section{Dissipatively coupled SSH (DSSH) model} \label{sec2}
Consider a 1-D lattice of sites A and B coupled dissipatively as depicted in Fig. \ref{sch}. This is analogous to the Hermitian SSH model, except for a notable difference in the off-diagonal elements, wherein, the real couplings $t_1$ and $t_2$ are now replaced by purely imaginary numbers leading to an effective non-Hermitian Hamiltonian given by
\begin{align}\label{eq5.1}
    H=-\sum_{i=1}^{N}\big((\Delta_1-i\Gamma_r)\ket{A_i}\bra{A_i}+(\Delta_2-i\Gamma_r)\ket{B_i}\bra{B_i}\big)\nonumber \\+\sum_{i=1}^{N}i\Gamma_1\ket{A_i}\bra{B_i}+\sum_{i=1}^{N-1}i\Gamma_2\ket{A_{{i+1}}}\bra{B_i}+h.c,
\end{align}
where $\Gamma_r=\gamma+\Gamma_1+\Gamma_2$ denotes the effective damping constant corresponding to $A_i$ and $B_i$ and we assume modes $A_i$ and $B_i$ decay at the same rate $\gamma$. The emergence of an effective Hamiltonian describing the above equation from considerations of a real-space Hermitian system will be explicated in subsequent sections. For simplicity, we set $\Delta_1=\Delta_2=0$. The effective Bloch Hamiltonian under PBC is provided by
\begin{align}\label{eq7}
    \mathcal{H}(k)=\begin{pmatrix}
-i\Gamma_r&i\Gamma_1+i\Gamma_2e^{-ik}\\
i\Gamma_1+i\Gamma_2e^{ik}&-i\Gamma_r
\end{pmatrix},
\end{align}
with eigenvalues $E_\pm=-i\Gamma_r\pm i\sqrt{\Gamma_1^2+\Gamma_2^2+2\Gamma_1\Gamma_2\cos{k}}$ where $k=\frac{2\pi n}{N}$, the parameters $\Gamma_j$, $j\in\{1,2\}$ are the absolute value of the strength of dissipative coupling and we have set the diagonal elements to be identical.  Owing to the non-Hermitian nature of the system, the right eigenvectors and their dual left eigenvector basis of $\mathcal{H}(k)$ are, in general, not identical. Let $\ket{R_\pm,k}$ and $\ket{L_\pm,k}$ be the bi-orthogonal right and left eigenvectors respectively of $\mathcal{H}(k)$ defined by
\begin{align}\label{eq8}
    \mathcal{H}(k)\ket{R_\pm,k}=E_\pm \ket{R_\pm,k} \nonumber\\
    \mathcal{H}^\dagger(k)\ket{L_\pm,k}=E_\pm^* \ket{L_\pm,k},
\end{align}
such that $\bra{L_m,k}\ket{R_n,k}=\delta_{m,n}$ where $m,n$ $\in \{+,-\}$ and $\delta_{m,n}$ is the Kronecker delta. Interestingly, $\mathcal{H}(k)$ is anti-Hermitian, i.e., $\mathcal{H}^\dagger(k)=-\mathcal{H}(k)$. As a result, the bi-orthogonal eigenvectors have the property
\begin{align}\label{eq9}
    \ket{R_\pm,k}=\ket{L_\pm,k}=\frac{1}{\sqrt{2}}\begin{pmatrix}
1\\
\pm ie^{i\phi(k)},
\end{pmatrix}
\end{align}
where $\phi(k)=-\arctan(\frac{\Gamma_1+\Gamma_2\cos{k}}{\Gamma_2\sin{k}})$. Subsequently, one can define the Berry connection involving the bi-orthogonal eigenvectors as $A_{\pm}(k)=i\bra{L_\pm,k}\partial_k\ket{R_\pm,k}$ and analogous to the Eq. (\ref{eq5}) of coherently coupled SSH model, the system is topological with $\nu_+=1$ for $|\Gamma_2|>|\Gamma_1|$. Note, \textit{en passant}, the constant diagonal decay is irrelevant for topological considerations. An interesting consequence of non-Hermiticity is the breakdown of BBC. In other words, the parameters corresponding to the energy gap closing in a non-Hermitian Bloch Hamiltonian do not, in general, signify the boundary between topological and trivial phases. On the contrary, owing to the anti-Hermitian nature of $\mathcal{H}(k)$, the dissipatively coupled system described in Fig. \ref{sch} follows BBC. In Fig. \ref{eigen} (b), we plot the eigenvalues of the system under PBC and OBC and when $\abs{\frac{\Gamma_1}{\Gamma_2}}<1$, it displays the conspicuous emergence of two distinct eigenvalues corresponding to the edge modes flanked on either side by the bulk modes. Not surprisingly, this is exactly the point of the vanishing energy gap between the Bloch modes corroborating BBC. 

\textit{Role of real diagonal terms in DSSH:}
 Consider now, the scenario $\Delta_1=-\Delta_2=\Delta$. The Bloch Hamiltonian under PBC modifies to
\begin{align}\label{eq9.1}
\mathcal{H}_{k}=\begin{pmatrix}
{\Delta}-i\Gamma_r & h(k) \\
-h^*(k) & -{\Delta}-i\Gamma_r 
\end{pmatrix},
\end{align}
where $h(k)=i\Gamma_1+i\Gamma_2e^{-ik}$. Observe that the effective momentum space Hamiltonian is anti-PT symmetric, in other words, $(\hat{PT})\mathcal{H}_{k}(\hat{PT})=-\mathcal{H}_{k}$. We may rewrite $h(k)=B_x-iB_y$, where $B_x=\Gamma_2\sin(k)$ and $B_y=-(\Gamma_1+\Gamma_2\cos(k))$ are pseudo magnetic fields providing an analogy to spin half particles interacting with magnetic fields. When discussing the topological properties of the system, the constant diagonal term $-i\Gamma_r$ is irrelevant. The eigenvalues of the system, ignoring $-i\Gamma_r$ are given by $E_{\pm}=\pm\sqrt{\bar{\Delta}^2-(B_x^2+B_y^2)}$ for $|\bar{\Delta}|>|h(k)|$ and $E_{\pm}=\pm i\sqrt{B_x^2+B_y^2-\bar{\Delta}^2}$ for $|\bar{\Delta}|<|h(k)|$. Here, we focus on the region where $|\bar{\Delta}|<|h(k)|$ and use the parametrization $\frac{\bar{\Delta}}{R}=\sinh(\theta)$, $\frac{B_x}{R}=\cosh(\theta)\cos(\phi)$, $\frac{B_y}{R}=\cosh(\theta)\sin(\phi)$, where $R=\sqrt{B_x^2+B_y^2-\bar{\Delta}^2}$ to obtain the right and left eigenvectors of the eigenvalue $\lambda_+$ as
\begin{align}\label{eq15}
\ket{R_+}=\frac{1}{\sqrt{2(1+i\sinh(\theta))}}\begin{pmatrix}
-i\cosh(\theta)e^{-i\phi}  \\
1+i\sinh(\theta) 
\end{pmatrix}
\end{align}
\begin{align}\label{eq16}
\ket{L_+}=\frac{1}{\sqrt{2(1+i\sinh(\theta))}}\begin{pmatrix}
-i\cosh(\theta)e^{-i\phi}  \\
1-i\sinh(\theta) 
\end{pmatrix}.
\end{align}

 The Berry connection of the system is defined as $A_+(k)=i\bra{L_+}\partial_k \ket{R_+}$. Utilizing the parametrization discussed above, we can recast the Berry connection as $A(k)=(\partial_k \phi) A(\phi)+(\partial_k \theta) A(\theta)$. If we consider, for example, a trajectory in the $\theta$-$\phi$ space, wherein, $\partial_k \theta=0$, the Berry connection is given by $A(k)=\frac{1}{2}(1-i\sinh(\theta)\partial_k\phi$. Note, when $\theta=0$, the equation for winding number 
\begin{align}\label{eq17}
\nu_+=\frac{1}{\pi}\oint A_+(k)dk,
\end{align}
 morphs into Eq. (\ref{eq5}) and the system is topological with $\nu_+=1$ for $|\Gamma_2|>|\Gamma_1|$. More precisely, the anti-PT symmetric system demonstrates partial topological phases for trajectories where $\partial_k \theta=0$, wherein, the real part of the winding numbers mimics the topology of a Hermitian SSH model. It is worth noting that the Hamiltonian $\mathcal{H}_{k}$ is not chiral symmetric, in other words $\sigma_z\mathcal{H}_{k}\sigma_z\neq-\mathcal{H}_{k}$.

In the following section, we provide two experimentally realizable protocols to engineer DSSH model.

\section{Realization of the dissipative SSH model}\label{sec3}
\begin{figure} \captionsetup{justification=raggedright,singlelinecheck=false}
 \centering
   \includegraphics[scale=1]{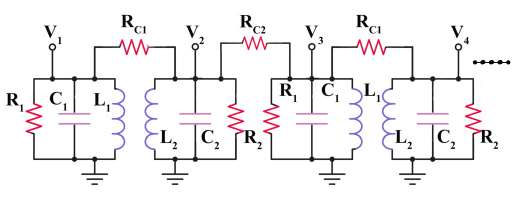}
\caption{Circuit model consisting of resistively coupled LCR resonators for the realization of dissipatively coupled SSH model resulting in the dynamics described by Eq. (\ref{eq18}).}
\label{circuit1}
\end{figure}

\textit{Circuit model:}
In this section, we provide a circuit model to construct the DSSH model employing an electrical circuit involving coupled amplifying LRC resonators connected in parallel through a coupling resistor as depicted in Fig. \ref{circuit1}. Upon solving for Kirchhoff's equations of motion for voltages, we obtain
\begin{align}\label{eq18}
\ddot{V}_{n}+\omega_{1}^2V_n+(\gamma_1+\Gamma_1+\Gamma_2)V_n=\Gamma_1\dot{\bar{V}}_n+\Gamma_2\dot{\bar{V}}_{n-1} \nonumber \\
\ddot{\bar{V}}_{n}+\omega_{2}^2\bar{V}_n+(\gamma_2+\Gamma_1+\Gamma_2)\bar{V}_n=\Gamma_1\dot{{V}}_n+\Gamma_2\dot{{V}}_{n+1}.
\end{align}
Here, $\omega_{1}=1/\sqrt{L_1C_1}$, $\omega_{2}=1/\sqrt{L_2C_2}$, $\Gamma_1=\frac{1}{R_{c1}C_1}$, $\Gamma_2=\frac{1}{R_{c2}C_2}$, $\gamma_i=\frac{1}{R_iC_i}$ and $L_{1,2}$, $R_{c}$, $C_{1,2}$ are respectively inductance, the two coupling resistances, capacitance of the constituent elements in the unit cell and we assume $C_1=C_2$. Note \textit{en passant}, the constants $\Gamma_1$ and $\Gamma_2$ represent the intra and inter-cellular couplings in the lattice. In the weak coupling and small detunings regime, that is, when $\{\Gamma_1,\Gamma_2\}<<\{\omega_1,\omega_2\}$ and $|\omega_1-\omega_2|<<\frac{\omega_1+\omega_2}{2}$, we can reduce the above equations by using the slowly varying envelope functions $v_{n}(t),\bar{v}_n(t)$, such that
\begin{align}\label{eq19}
V_n(t)=\frac{v_{n}(t)e^{-i\omega_0(t)}+c.c}{2} \nonumber \\
\bar{V}_n(t)=\frac{\bar{v}_{n}(t)e^{-i\omega_0(t)}+c.c}{2},
\end{align}
where $\omega_0=\frac{\omega_1+\omega_2}{2}$. Employing Eq. (\ref{eq19}), the dynamics of the envelope functions are obtained as 
\begin{align}\label{eq20}
\dot{v}_n=-i\frac{\bar{\Delta}-i(\gamma_1+\Gamma_1+\Gamma_2)}{2}v_n+\frac{\Gamma_1}{2}\bar{v}_n+\frac{\Gamma_2}{2}\bar{v}_{n-1} \nonumber \\
\dot{\bar{v}}_n=i\frac{\bar{\Delta}+i(\gamma_2+\Gamma_1+\Gamma_2)}{2}\bar{v}_n+\frac{\Gamma_1}{2}{v}_n+\frac{\Gamma_2}{2}{v}_{n+1},
\end{align}
with $\bar{\Delta}=\frac{\omega_1-\omega_2}{2}$. We assume propagating solutions for the sub-lattice elements, that is
\begin{align}\label{eq21}
v_n=\sum_k v_{k,\omega}e^{i(kn-\omega t)}+c.c \nonumber \\
\bar{v}_n=\sum_k \bar{v}_{k,\omega}e^{i(kn-\omega t)}+c.c,
\end{align}
where $k$ is the wave vector and the lattice constant is taken to be unit. Substituting Eq. (\ref{eq21}) into Eq. (\ref{eq20}), we arrive at the eigenvalue equation $(H_{k}-\omega)X=0$, where $X^{T}=$[ $v_{k,\omega}$ $\bar{v}_{k,\omega}$] and
\begin{align}\label{eq22}
H_{k}=\begin{pmatrix}
\bar{\Delta}-i(\gamma_1+\Gamma_1+\Gamma_2) & i(\frac{\Gamma_1}{2}+\frac{\Gamma_2}{2}e^{-ik}) \\
i(\frac{\Gamma_1}{2}+\frac{\Gamma_2}{2}e^{-ik}) & -\bar{\Delta}-i(\gamma_2+\Gamma_1+\Gamma_2) 
\end{pmatrix}.
\end{align}
The Hamiltonian $H_k$ is equivalent to $\mathcal{H}_{k}$. In other words, the circuit lattice is topologically equivalent to a dissipatively coupled SSH model.

\begin{figure}
\captionsetup{justification=raggedright,singlelinecheck=false}
 \centering
   \includegraphics[scale=1.25]{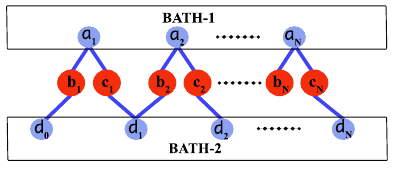}
\caption{Schematic of the coupled oscillator system described by the Hamiltonian in Eq. (\ref{eq10}) comprising of a bath of oscillators coupled with a system of otherwise non-interacting bosonic modes.}
\label{sch1}
\end{figure}

\textit{Photonic systems:}
We begin by considering the following generic Hamiltonian comprising of a chain of coherently coupled bosonic sub-lattice elements $a_i$, $b_i$, $c_i$ and $d_i$ under OBC as depicted in Fig. \ref{sch1}. 
\begin{align}\label{eq10}
\mathcal{H}/\hbar=\sum_{i=1}^{N}\omega_{b,i} b_i^{\dagger}b_i+\sum_{i=1}^{N}\omega_{c,i} c_i^{\dagger}c_i +\sum_{i=1}^{N}\omega_{a,i} a_i^{\dagger}a_i+\sum_{i=0}^{N}\omega_{d,i} d_i^{\dagger}d_i \nonumber \\
+\sum_{i=1}^{N} [g_1b_i^{\dagger}(a_{i}+d_{i-1})+h.c]+\sum_{i=1}^{N} [g_2c_i^{\dagger}(a_{i}+d_{i})+h.c]
\end{align} 
Here, $\omega_{x,i}$ characterizes the resonance frequencies of the modes $x_i$, where $x\in \{a,b,c,d\}$ and $g_1$, $g_2$ $\in \mathbb{R}$ are the strength of dispersive coupling between the modes. 
We assume that all the modes $b_i$ and $c_i$ decay at approximately the same rate $\gamma$ whereas the modes $a_i$ and $d_i$ decay at rates $\kappa_1$ and $\kappa_2$ respectively. Further, we set $\Delta_{b,i}=\Delta_1=\omega_{b,1}-\omega_{a,1}$, $\Delta_{c,i}=\Delta_2=\omega_{c,1}-\omega_{a,1}$, $\omega_{a,i}=\omega_{b,i}=\omega_{a,1}$ and $\kappa_1>\kappa_2$. In the weak coupling domain, that is, when the leakage rates $\kappa_i$ strongly dominates the dynamics of the system, in other words, $\{g_1, g_2, \gamma, \Delta_{b,1}, \Delta_{c,1}\}<<\{\kappa_1,\kappa_2\}$, we can adiabatically eliminate the $a_i$ and $d_i$ modes resulting in the effective momentum space Hamiltonian of the system under PBC ($d_0=d_{N}$) in the frame rotating at $(\Delta_1+\Delta_2)/2$ as 
\begin{align}\label{eq14}
\mathcal{H}_{k}=\begin{pmatrix}
\bar{\Delta}-i\Gamma_r & h(k) \\
-h^{*}(k) & -\bar{\Delta}-i\Gamma_r 
\end{pmatrix},
\end{align}
where, we set $g_1=-g_2=g$, $\bar{\Delta}=(\Delta_1-\Delta_2)/2$, $\Gamma_r=\gamma+\Gamma_1+\Gamma_2$, $h(k)=i\Gamma_1+i\Gamma_2e^{-ik}$ and $k$ is the lattice constant equivalent to Eq. (\ref{eq9.1}) and $\Gamma_i=\frac{g^2}{\kappa_i}$. Note that the pair of modes $b_i$ and $c_{i}$ form a unit cell with intra and inter cell couplings $i\frac{g^2}{\kappa_1}$ and $i\frac{g^2}{\kappa_2}$ respectively equivalent to the system in Fig. \ref{sch} under dissipative settings with $\Gamma_i$ replaced by $\frac{g^2}{\kappa_i}$.

\section{Non-reciprocity, phase-dependent topological transitions and skin effect}\label{sec4}
In the previous section, we briefly mentioned the breakdown of BBC in non-Hermitian systems. In particular, non-reciprocal (chiral) coupling between sub-lattice elements under OBC culminates in the skin effect which is the exponential localization of right and left eigenvectors at the lattice boundaries without any distinction between bulk and edge modes. In addition, the points in the parametric space corresponding to the closing of energy gap under PBC do not indicate the emergence of eigenmodes with eigenvalues $-i\Gamma_r$ under OBC requiring a bi-orthogonal modification of the BBC. In the following, we discuss the construction of non-reciprocal couplings in DSSH lattice leading to phase-dependent topological transitions and skin effect. 

Consider now, a one-dimensional lattice of bosonic modes $b$, $c$ coupled with auxiliary modes $a$ as depicted in Fig. \ref{sch1} where we have now switched on the coherent coupling between $b_i$ and $c_i$ modes. The system is characterized by the Hamiltonian
\begin{align}\label{eq23}
    \tilde{\mathcal{H}}/\hbar=\mathcal{H}/\hbar+\sum_{i=1}^{N} [Gb_i^{\dagger}c_i+h.c],
\end{align}
where $\mathcal{H}$ is given by Eq. (\ref{eq10}) and $G=|G|e^{i\alpha}$. Once again, in the weak coupling domain, that is, when $\{g_1,g_2,|G|,\gamma\}$ are significantly less than the cavity leakage $\kappa_i$ and setting $\Delta_{b,i}=\Delta_1=\omega_{b,1}-\omega_{a,1}$, $\Delta_{c,i}=\Delta_2=\omega_{c,1}-\omega_{a,1}$, $\omega_{a,i}=\omega_{b,i}=\omega_{a,1}$, $\kappa_1>\kappa_2$ and $g_1=-g_2=g$, we can obtain an effective system between modes $b_i$ and $c_i$. The system under PBC translates into the following Bloch Hamiltonian in the frame rotating at $(\Delta_1+\Delta_2)/2$
\begin{align}\label{eq29}
\mathcal{H}_{k}=\begin{pmatrix}
\bar{\Delta}-i\Gamma_r & i\Gamma_-+i\Gamma_2e^{-ik} \\
i\Gamma_++i\Gamma_2e^{ik} & -\bar{\Delta}-i\Gamma_r 
\end{pmatrix},
\end{align}
where $\Gamma_\pm=\Gamma_1\mp |G|\sin{\alpha}-i|G|\cos{\alpha}$, $\bar{\Delta}=(\Delta_1-\Delta_2)/2$ and $\Gamma_r=\gamma+\Gamma_1+\Gamma_2$. Notice, the conspicuous emergence of a purely dissipative form of non-reciprocal coupling between the subsystems when $\alpha=\pi/2$.

Note that the non-Hermitian system described by the aforementioned Hamiltonian does not follow BBC. To elucidate this in detail, let us begin by considering the system under OBC. Before expounding the analysis of the full system, it is worthwhile to explicate the properties of the system in the absence of $c_N$ and to simplify the analysis, we set $\bar{\Delta}=0$ for the remaining part of this section. When the lattice terminates in $b_N$, the $2N-1$ dimensional Hamiltonian of the system $H_{broken}^b$ supports bi-orthogonal eigenstates with eigenvalue $-i\Gamma_r$ of the form \cite{PhysRevLett.121.026808, PhysRevB.97.241405}
\begin{align}\label{eq30}
\ket{R}^b={N}_R^b\sum_{n=0}^{N-1}\Big(-\frac{\Gamma_2}{\Gamma_+}\Big)^{N-n}b_{n+1}^{\dagger}\ket{0} \nonumber \\
\ket{L}^b={N}_L^b\sum_{n=0}^{N-1}\Big(-\frac{\Gamma_2}{\Gamma_-^*}\Big)^{N-n}b_{n+1}^{\dagger}\ket{0},
\end{align} 
where $N_R$, $N_L$ are normalization constants such that $\bra{L}\ket{R}=1$ provided by $N_L^{b*}N_R^b=Z^{N+1}\frac{(Z^{-1}-1)}{1-Z^N}$, $Z=\frac{\Gamma_-\Gamma_+}{\Gamma_2^2}$ and
\begin{align}\label{eq30.1}
    H_{broken}^b=\begin{bmatrix} 
    -i\Gamma_r & i\Gamma_- & 0 & 0 &\dots &0 \\
    i\Gamma_+ & -i\Gamma_r & i\Gamma_2 & 0 & \dots & 0 \\
    0 & i\Gamma_2 & -i\Gamma_r & i\Gamma_-& \dots &0 \\
     0 & 0 & i\Gamma_+ & -i\Gamma_r& \dots & 0 \\
    0 & 0 & \dots & \dots & \dots & i\Gamma_2 \\
    0 & 0 & \dots & \dots& i\Gamma_2 & -i\Gamma_r 
    \end{bmatrix}_{2N-1}.
\end{align}
This is due to destructive interference at $c_i$ sites and observe that $\ket{R}^b$ and $\ket{L}^b$ can be written as a column matrix, for instance,
\begin{align}\label{eq30.2}
    \ket{R}^b=\begin{bmatrix} 
    \Big(-\frac{\Gamma_2}{\Gamma_+}\Big)^N & \\
    0  \\
    \Big(-\frac{\Gamma_2}{\Gamma_+}\Big)^{N-1} \\
     \vdots \\
    0  \\
    -\frac{\Gamma_2}{\Gamma_+} 
    \end{bmatrix}_{2N-1},
\end{align}
and clearly $ H_{broken}^b\ket{R}^b=-i\Gamma_r\ket{R}^b$. Notice that when $\alpha=\pi/2$, Eq. (\ref{eq30}) morphs into
\begin{align}\label{eq31}
\ket{R}^b={N}_R^b\sum_{n=0}^{N-1}\Big(-\frac{\Gamma_2}{\Gamma_1+\abs{G}}\Big)^{N-n}b_{n+1}^{\dagger}\ket{0} \nonumber \\
\ket{L}^b={N}_L^b\sum_{n=0}^{N-1}\Big(-\frac{\Gamma_2}{\Gamma_1-\abs{G}}\Big)^{N-n}b_{n+1}^{\dagger}\ket{0},
\end{align}
which is, identical to the results for a coherently coupled SSH model with non-reciprocal intra-cell couplings. By the same token, one can construct eigenstates of complex energy $-i\Gamma_r$ of the Hamiltonian $H_{broken}^c$ of the lattice which is broken at the other end, i.e., when the lattice ends on either side with $c_i$ sites as
\begin{align}\label{eq31.0}
    \ket{R}^c={N}_R^c\sum_{n=1}^{N}\Big(-\frac{\Gamma_2}{\Gamma_-}\Big)^{N}c_{n}^{\dagger}\ket{0} \nonumber \\
\ket{L}^c={N}_L^c\sum_{n=1}^{N}\Big(-\frac{\Gamma_2}{\Gamma_+^*}\Big)^{N}c_{n}^{\dagger}\ket{0},
\end{align}
where $N_L^{c*}N_R^c=N_L^{b*}N_R^b=Z^{N+1}\frac{(Z^{-1}-1)}{1-Z^N}$.

In stark contrast to Hermitian systems where the absolute value square of the coeffcient of the column matrix in Eq. (\ref{eq30.2}) would represent the probability of finding the excitation at the $n^{th}$ unit cell, non-Hermitian systems necessitate a bi-orthogonally defined projection to the $n^{th}$ unit cell. Subsequently, one can define a biorthogonal projection operator $P_n$ to the $n^{th}$ unit cell of the lattice as $P_n=\ket{b,n}\bra{b,n}+\ket{c,n}\bra{c,n}$ where $\ket{b,n}=b_n^\dagger\ket{0}$ and $\ket{c,n}=c_n^\dagger\ket{0}$. For example, projecting the states in Eq. (\ref{eq30}) on to the $n^{th}$ unit cell provides
\begin{align}\label{eq31.1}
    \bra{L}^bP_n\ket{R}^b=\frac{Z^{n+1}(Z^{-1}-1)}{1-Z^N}.
\end{align}
It is, therefore, apparent that for $|Z|<1$, the excitation is exponentially localized at the left edge ($n=1$) whereas $|Z|>1$ localizes the state at the right edge $n=N$. Similarly, when the lattice terminates with a $c_i$ mode on either side ($b_1$ is absent), one can obtain analogous results with the excitation localized at the right (left) edge for $|Z|<1 (|Z|>1)$ due to mirror symmetry. 

The results of the broken chain system can now be used to extract the physics of the full system in the thermodynamic limit (large $N$). Consider the Hamiltonian of the full system
\begin{figure}
\captionsetup{justification=raggedright,singlelinecheck=false}
 \centering
   \includegraphics[scale=0.93]{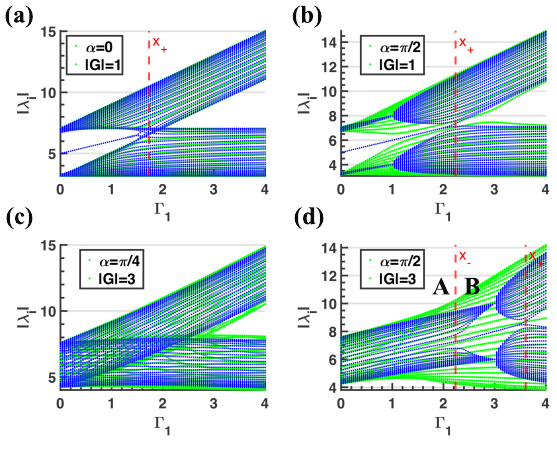}
\caption{The absolute value of eigenvalues of $H_{full}$ under OBC (blue) and PBC (green) for different values of $\alpha$ and $|G|$ with $\Gamma_2=2$, $\gamma=3$, the number of units cells $N=25$. The vertical lines represent the points $x_\pm$ obtained from Eq. (\ref{eq33}) as $x_\pm=\sqrt{A_\pm}$ for non-negative values of $A_\pm$. The two states with complex energy $-i\Gamma_r=-i(\gamma+\Gamma_1+\Gamma_2)$ appear in the region $\Gamma_1<x_-$ and $\Gamma_1<x_+$ as isolated blue lines in the middle in (a), (b), whereas they appear in the region $x_-<\Gamma_1<x_+$ in (d).}
\label{energy}
\end{figure}
\begin{align}\label{eq31.2}
    H_{full}=\begin{bmatrix} 
    -i\Gamma_r & i\Gamma_- & 0 & 0 &\dots &0 &0\\
    i\Gamma_+ & -i\Gamma_r & i\Gamma_2 & 0 & \dots & 0 &0\\
    0 & i\Gamma_2 & -i\Gamma_r & i\Gamma_-& \dots &0 &0\\
     0 & 0 & i\Gamma_+ & -i\Gamma_r& \dots & 0 & 0\\
    0 & 0 & \dots & \dots & \dots & i\Gamma_2 &0\\
    0 & 0 & \dots & \dots& i\Gamma_2 & -i\Gamma_r & i\Gamma_- \\
    0 & 0 & \dots & \dots& 0 & i\Gamma_+& -i\Gamma_r
    \end{bmatrix}_{2N},
\end{align}
and states 
\begin{align}\label{eq31.3}
   \ket{\psi}_R^{\pm}=\frac{1}{\sqrt{2}}\Biggl\{\begin{bmatrix} 
    \ket{R}^b \\
    0
    \end{bmatrix}\pm \begin{bmatrix} 
    0 \\
    \ket{R}^c
    \end{bmatrix}\Biggl\}\nonumber \\
    \ket{\psi}_L^{\pm}=\frac{1}{\sqrt{2}}\Biggl\{\begin{bmatrix} 
    \ket{L}^b \\
    0
    \end{bmatrix}\pm \begin{bmatrix} 
    0 \\
    \ket{L}^c
    \end{bmatrix}\Biggl\}.
\end{align}
\begin{figure} \captionsetup{justification=raggedright,singlelinecheck=false}
 \centering
   \includegraphics[scale=1.1]{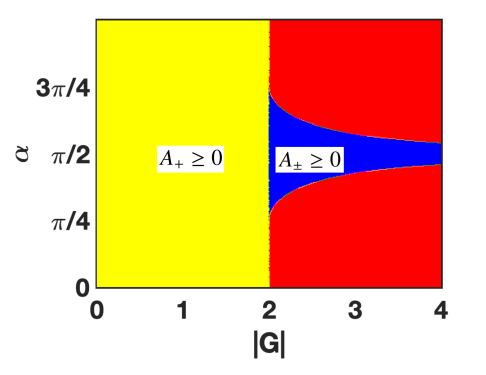}
\caption{The phase diagram of the system as a function of $\alpha$ and $|G|$ for $\Gamma_2=2$ characterized by the real positive values of $A_\pm$ demarcating the topological boundaries. In particular, the region in red depicts the topologically trivial parametric domain.}
\label{phase}
\end{figure}
Clearly, $\bra{\psi}_L^{\pm} \ket{\psi}_R^{\pm}=1$ and $\bra{\psi}_L^{\mp} \ket{\psi}_R^{\pm}=0$ and it is straightforward to obtain 
\begin{align}\label{eq31.4}
    H_{full} \ket{\psi}_R^{\pm}= \begin{bmatrix} 
    \mp i\Gamma_2 N_R^c & \\
    0  \\
     \vdots \\
    0  \\
    -i\Gamma_2 N_R^b
    \end{bmatrix}_{2N}-i\Gamma_r\ket{\psi}_R^{\pm} \nonumber \\
    H_{full}^\dagger \ket{\psi}_L^{\pm}=\begin{bmatrix} 
\pm i\Gamma_2 N_L^c & \\
    0  \\
     \vdots \\
    0  \\
    i\Gamma_2 N_L^b
    \end{bmatrix}_{2N}+i\Gamma_r\ket{\psi}_L^{\pm},
\end{align}
where we have defined $N_R^b=N_R^c=\Big(\frac{Z^{N+1}(Z^{-1}-1)}{1-Z^N}\Big)^{1/2}$ and $N_L^b=N_L^c=N_R^{b*}$.
It is conspicuous from the above expression that $\ket{\psi}_{R,L}^{\pm}$ represent bi-orthogonal eigenstates of $H_{full}$ with complex energy $-i\Gamma_r$ for large $N$ for $N_R^b\rightarrow 0$, \textit{viz}, if $|Z|<1$ (and not for $|Z|\geq 1$) as the normalization factors in the first part of the RHS of Eq. (\ref{eq31.4}) approach zero. In other words, the states in Eq. (\ref{eq31.3}) are the eigenstates of $H_{full}$ with eigenvalue $-i\Gamma_r$ ($\Gamma_r$-modes) for
\begin{align}\label{eq32}
    \sqrt{(\Gamma_1^2-\abs{G}^2)^2+4\abs{G}^2\Gamma_1^2\cos^2{\alpha}}<\Gamma_2^2.
\end{align}
It is worth noting that for $\alpha=\pi/2$, the condition for bi-orthogonal edge modes modifies to $\Gamma_1^2-\abs{G}^2<\Gamma_2^2$. The Eq. (\ref{eq32}) may be rewritten as $(\Gamma_1^2-A_+)(\Gamma_1^2-A_-)<0$, where
\begin{align}\label{eq33}
    A_{\pm}=-|G|^2\cos{(2\alpha)}\pm\sqrt{\Gamma_2^4-|G|^4\sin^2{(2\alpha)}}
\end{align}
for real values of the RHS of Eq. (\ref{eq33}). Therefore, the $\Gamma_r$-modes of the system under OBC occur in the region where $\Gamma_1^2<A_+$ and $\Gamma_1^2>A_-$. It is worth noting that the system does not incur $\Gamma_r$-modes for $\Gamma_2^4-|G|^4\sin^2{(2\alpha)}<0$. In Fig. \ref{energy}, we plot the absolute value of eigenvalues of the full system under OBC and PBC for different values of $\alpha$ and $|G|$ for $\Gamma_2=2$ where we have defined $x_\pm=\sqrt{A_\pm}$ for non-negative values of $A_\pm$. In Fig. \ref{energy} (a-b), for $|G|=1$, we observe that $A_+>0$ and $A_-<0$ lending $\Gamma_r$-modes for $|\Gamma_1|<x_+$ depicted by the two isolated blue lines. In stark contrast, for $|G|=3$ and $\alpha=\pi/4$, $\Gamma_2^4-|G|^4\sin^2{(2\alpha)}<0$, leading to the conspicuous absence of $\Gamma_r$-modes as depicted in Fig. \ref{energy} (c). However, $\alpha=\pi/2$, $|G|=3$ (Fig. \ref{energy} (d)) renders $A_\pm>0$ which affords $\Gamma_r$-modes on either in the region $x_-<\Gamma_1<x_+$, demonstrating phase dependent nature of topological transitions. Note that the green curves in Fig. \ref{energy} correspond to the absolute value of energy under OBC. Clearly, the points where the blue curves approach $-i\Gamma_r$ do not match with that of the green curves as a consequence of the breakdown of BBC. In Fig. \ref{phase}, we plot the phase (not to be confused with $\alpha$) diagram of the system as a function of $\alpha$ and $|G|$ depending on the real positive values of $A_\pm$, clearly demarcating the topological boundaries. For the region depicted in yellow, we have only the $A_+\geq 0$, begetting $\Gamma_r$-modes modes for $|\Gamma_1|<\sqrt{A_+}$ as displayed in Fig. \ref{energy} (a-b). The region in blue, however, provides $A_\pm\geq 0$, lending $\Gamma_r$-modes when $\Gamma_1^2<A_+$ and $\Gamma_1^2>A_-$ as demonstrated in Fig. \ref{energy} (d). In contrast, the region in red prohibits real positive values for $A_\pm$ and therefore does not lend itself to a topological description evident from Fig. \ref{energy} (c).

\begin{figure} \captionsetup{justification=raggedright,singlelinecheck=false}
 \centering
   \includegraphics[scale=0.92]{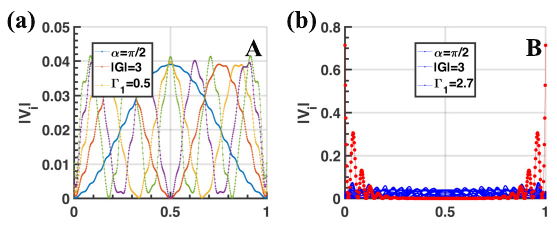}
\caption{(a) The absolute value of $N$ components (equally spaced between 0 and 1) of the vector $V^i$ for $i=1$ to $5$; (b) absolute value of $N$ components of $V^i$ for $i=1$ to $N$. Note that the red curves in (b) represent the two topological edge-modes of the system (concentrated at both the right and left edges) and we have set the number of unit cells $N=25$, $\Gamma_2=2$, $\gamma=3$. All the $N$ vectors $V^i$ are bulk-modes for $\alpha=\pi/2$, $|G|=3$ and $\Gamma_1=0.5$ and for clarity, we only plot five of them in (a) as the rest of them have similar behavior.}
\label{proj}
\end{figure}

To provide more substance to the above discussion, we plot, in Fig. \ref{proj} (a-b) the absolute value of $N$ components (equally spaced between 0 and 1) of the vector $V^i=$[ $\pi_1^i$ $\pi_2^i$ ... $\pi_N^i$ ] for two different regions of Fig. \ref{energy}(d) with $i\in\{1,2,..N\}$ and 
\begin{align}\label{eq33.1}
    \pi_n^i=\frac{\bra{L_i^{full}}P_n\ket{R_i^{full}}}{\bra{L_i^{full}}\ket{R_i^{full}}}
\end{align}
\begin{figure} \captionsetup{justification=raggedright,singlelinecheck=false}
 \centering
   \includegraphics[scale=0.93]{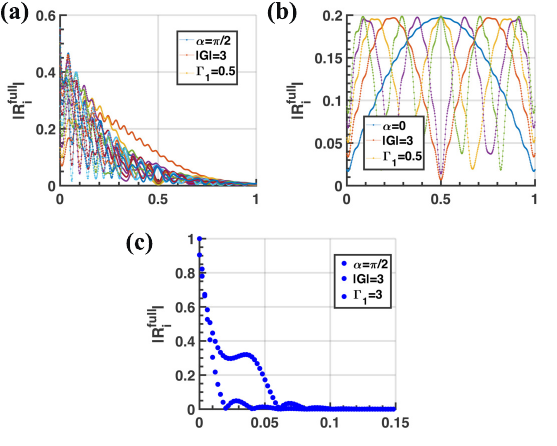}
\caption{ (a) The absolute value of the eigenvectors (equally spaced between 0 and 1) $\ket{R_i^{full}}$ for $\alpha=\pi/2$, $|G|=3$, $\Gamma_1=0.5$, clearly demonstrating the eigenmodes, all of which are concentrated on the left edge of the system; (b) absolute value of the eigenvectors $\ket{R_i^{full}}$ for $\alpha=0$, $|G|=3$, $\Gamma_1=0.5$ with no non-reciprocity in couplings leading to the absence of edge effects; (c) plot of the absolute value of the eigenvectors $\ket{R_i^{full}}$ (for the region between zero and 0.15) with $\alpha=\pi/2$, $|G|=3$, $\Gamma_1=3$ with $\Gamma_+\rightarrow 0$ giving rise to extreme non-reciprocity and amplification of edge population when compared to (a). The quantity $\abs{R_i^{full}}$ is negligibly small beyond $0.15$ in (c) and analogous to Fig. \ref{proj} (a), for clarity, we have only plotted $V^i$ for $i=1$ to $5$ in (b). The number of unit cells $N=25$, $\Gamma_2=2$, $\gamma=3$.}
\label{reig}
\end{figure}where $\ket{R_i^{full}}$, $\ket{L_i^{full}}$ are the right and eigenvectors of $H_{full}$ respectively. The Fig. \ref{proj} (a) corresponds to a point in the topologically trivial region $A$ of the Fig. \ref{energy}(d), that is, for $\alpha=\pi/2$, $\Gamma_1=0.5$, $|G|=3$ hallmarked by the absence of edge modes. In contrast, the Fig. \ref{proj} (b) depicts the absolute value of the projection for $\alpha=\pi/2$, $\Gamma_1=2.7$, $|G|=3$, a point in the region $B$ of Fig. \ref{energy}(d). Not surprisingly, edge states (in red) springs into existence and they correspond to Eq. (\ref{eq31.3}) matching the two isolated central modes of Fig. \ref{energy}(d) possessing complex energy $-i\Gamma_r$. To be precise, the two edge-modes in Fig. \ref{proj} (b) correspond to $V^i$ where $\ket{R_i^{full}}$,  $\ket{L_i^{full}}$ are replaced, respectively, by the states $\ket{\psi}_R^{\pm}$ and $\ket{\psi}_L^{\pm}$. On the contrary, the absolute value of the components of eigenvectors $\ket{R_i^{full}}$ and $\ket{L_i^{full}}$ would provide starkly dissimilar results with all the eigenstates concentrated at the boundaries, otherwise known as the skin effect. Skin effect, \textit{viz}, the unusual localization of a large number of eigenstates at the boundaries \cite{PhysRevLett.116.133903}, a consequence of non-reciprocal coupling, leads to pronounced sensitivity of the bulk to boundary conditions. In Fig. \ref{reig}(a), we plot $\abs{R_i^{full}}$, the absolute value of the $N$ components of the column matrix (equally spaced between 0 and 1) representing $\ket{R_i^{full}}$ for $\alpha=\pi/2$ clearly demonstrating the accumulation of eigenstates at the boundaries of the system owing to the non-reciprocal nature of dissipative coupling. On the contrary, $\alpha=0$ does not incur any non-reciprocity in coupling, culminating in the absence of skin effect as depicted in Fig. \ref{reig} (b). In stark contrast to Fig. \ref{reig} (a, b), the condition $\alpha=\pi/2$ and $|G|=\Gamma_1$ results in $\Gamma_+\rightarrow0$, i.e., extreme non-reciprocity and skin-effect. This is manifested in Fig. \ref{reig} (c), showcasing the remarkably high localization of the right eigenvectors at the left edge.
\section{conclusions}\label{sec5}
In conclusion, we considered SSH models with a dissipative form of coupling between the subsystems and discussed some of the interesting physics ensuing from such models. In particular, we provided two distinct schemes for the realization of DSSH models in the context of bosonic systems and electrical LCR resonators. We showed that a collection of resistively coupled LCR resonators mimic the topology of DSSH models by solving the Kirchhoff's equation for voltages. In the framework of bosonic systems, we observed that a system of non-interacting oscillators interacting with an engineered bath of modes possessing considerably small lifetimes compared to other system parameters is equivalent to a DSSH model. Further, by enabling the coherent interaction between the oscillators under consideration, we showed that the system affords non-reciprocal dissipative couplings eliciting topological transitions governed by the phase of the coherent interaction strength and skin effect. Note that our analyses are generic, relevant to a large class of systems, especially in microwave to optical settings and merits immediate realization in the experiments. 
\section{Acknowledgements}
GSA and MOS acknowledge the support of the Air Force Office of Scientific Research [AFOSR award no FA9550-20-1-0366 DEF] and the Robert A. Welch Foundation (Grant No. A-1261, Grant no A-1943). JMPN acknowledges the support of the Herman F. Heep and Minnie Belle Heep Texas A\&M University endowed fund held/administered by the Texas A\&M Foundation.

\appendix
\section{Kirchhof's equations for the circuit DSSH model}
\label{Appendix:a}
\begin{figure} \captionsetup{justification=raggedright,singlelinecheck=false}
 \centering
   \includegraphics[scale=1.1]{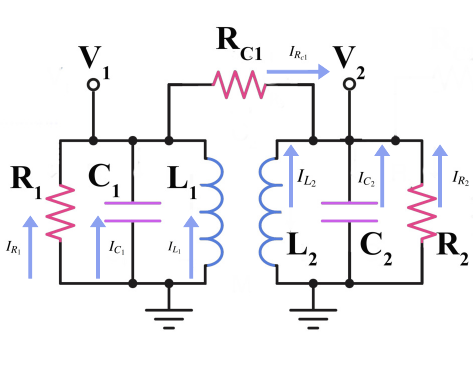}
\caption{The unit cell of the circuit SSH model described in Fig. \ref{circuit1}. The blue arrows represent the chosen direction currents in the system.}
\label{circuit2}
\end{figure}

We begin by considering the two blocks of LCR circuits, in other words, a dimer, coupled through a resistor as depicted in Fig. \ref{circuit2}.  After the choice of direction of currents as illustrated in the figure, we use the well-known Kirchoff's circuit laws to explicate the dimer dynamics. 
\begin{align}\label{eq.a1}
I_{R_{c1}}=I_{L_1}+I_{C_1}+I_{R_1}=-(I_{L_2}+I_{C_2}+I_{R_2}),
\end{align}
\begin{align}\label{eq.a2}
V_n=-L_n\frac{dI_{L_n}}{dt}=-\frac{1}{C_n}\int_{0}^{t}I_{C_n}(t^{\prime})dt^{\prime}=-I_{R_1}R_1,
\end{align}
\begin{align}\label{eq.a3}
V_1-V_2=I_{R_{c1}}R_{c1},
\end{align}
where n=1,2. Employing Eq. (\ref{eq.a2}) and Eq. (\ref{eq.a3}) into Eq. (\ref{eq.a1}), we obtain
\begin{align}\label{eq.a4}
\ddot{V_i}+\frac{1}{C_i}\Big(\frac{1}{R_{c1}}+\frac{1}{R_i}\Big)\dot{V_i}+\frac{V_i}{L_iC_i}=\frac{\dot{V_j}}{R_{c1}C_i}, \quad i\neq j.
\end{align}
Upon redefining $\omega_i=\frac{1}{\sqrt{L_iC_i}}$, $\Gamma_i=\frac{1}{R_{c1}C_i}$ and $\gamma_i=\frac{1}{R_iC_i}$ Eq. (\ref{eq.a4}) reduces to 
\begin{align}\label{eq.a5}
\ddot{V}_i+(\gamma_i+\Gamma_i)V_i+\omega_i^2V_i=\Gamma_i\dot{V_j}, \quad i\neq j,
\end{align}
The Eq. (\ref{eq.a5}) can be further simplified if we assume $V_{i}(t)=\frac{1}{2}u_i(t)e^{-i\omega_0t}+c.c$, where $\omega_{0}=\frac{1}{2}(\omega_1+\omega_2)$ and $u_i(t)$ is a slowly varying envelope. In addition, we assume that $k<<\omega_i$, $\omega_1$ close to $\omega_2$ and $C_1=C_2$. Under these conditions, the Eq. (\ref{eq.a5}) can be approximated to 
\begin{align}\label{eq.a6}
\begin{pmatrix}
\dot{u_1}\\
\dot{u_2}
\end{pmatrix}=-i\frac{1}{2}\begin{pmatrix}
\omega_1-\omega_2-i(\gamma_1+\Gamma_1)&i\Gamma_1\\
i\Gamma_1&\omega_2-\omega_1-i(\gamma_2+\Gamma_1)
\end{pmatrix}\begin{pmatrix}
{u_1}\\
{u_2}
\end{pmatrix}.
\end{align}
Extending the analysis to the full system in Fig. \ref{circuit1}, we obtain Eq. (\ref{eq18}).

\bibliography{references}
\end{document}